\documentclass[twocolumn,showpacs,groupedaddress]{revtex4}

\usepackage{graphicx}

\begin{document}

\newcommand{\ket}[1]{|#1\rangle}
\newcommand{\bra}[1]{\langle#1|}
\newcommand{\langlep}[1]{_p\langle}

\title{
Indirect quantum control for finite-dimensional coupled systems}
\author{ J. Nie$^1$, H. C. Fu$^2$, X. X. Yi$^1$}

\affiliation{$^1$School of Physics and Optoelectronic Technology,
Dalian University of Technology, Dalian 116024, China \\
$^2$School of Physics Science and Technology,
Shenzhen University, Shenzhen 518060, China}

\date{\today}

\begin{abstract}
We present a new analysis on the quantum  control for a quantum
system coupled to a quantum probe. This analysis is based on the
coherent control for the quantum system  and a hyperthesis that the
probe can be prepared in  specified  initial states. The results
show that a quantum system can be manipulated by probe
state-dependent coherent control. In this sense, the present
analysis  provides a new control scheme which combines the coherent
control and state preparation technology.
\end{abstract}

\def\aL{ \hat{a}^{\dag}_{L} }
\def\aR{ \hat{a}^{\dag}_{R} }
\def\bL{ \hat{b}^{\dag}_{L} }
\def\bR{ \hat{b}^{\dag}_{R} }

\pacs{ 03.65.-w, 03.67.Mn,02.30.Yy } \maketitle

\section{Introduction}

Controlling the time evolution
\cite{rabitz00,lloyd00,jurdjevic97,butkovskiy00,
blaquiere87,ramakrishna95,schirmer01,fu01} of a quantum system is a
major task required for quantum information processing. Several
approaches to the control of a quantum system have been proposed in
the past decade, which can be divided into coherent (unitary) and
incoherent  (non-unitary) control, according to how the controls
enter the dynamics. In the coherent control scheme, the controls
enter the dynamics through the system Hamiltonian. It affects the
time evolution of the system state, but not its spectrum, i.e., the
eigenvalues of the target density matrix $\rho_f$ remain unchanged
in the dynamics,  due to the unitarity of the evolution. In the
incoherent control scheme \cite{romano06,romano06l,xue06,fu07}, an
auxiliary system, called probe, is introduced to manipulate the
target system through their mutual interaction. This incoherent
control scheme is of relevance whenever the system dynamics can not
be directly accessed, and it provides a non-unitary  evolution which
is capable for transferring all initial states (pure or mixed) into
an arbitrary pure or mixed state. This breaks the limitation for the
coherent control mentioned above.

To be specific, Romano and colleagues \cite{romano06,romano06l} have
investigated  accessibility and controllability of a quantum bit (or
a two-level system) coupled to a quantum probe, under the condition
that  the external control affects only the probe. This analysis is
based on the Cartan decomposition \cite{helgason78,zhang03,
cabrera07} of the dynamics, and hence it is involved for
high-dimensional systems. In Ref.\cite{pechen06} the authors propose
a leaning control with a non-equilibrium environment. The results
show that by tailoring the dissipative dynamics we can control the
quantum system form a given state to a limited set of
states (reachable states).

In this paper, we first examine the controllability for
two-dimensional systems, then extend the approach to
finite-dimensional quantum systems. This analysis is based on the
quantum coherent control scheme and the hypothesis that the probe
can be prepared in a specified  initial state.  The advantages of
this scheme are threefold. Firstly, it overcomes the difficulty  of
the Cartan decomposition   based analysis. Secondly, it brings a
connection between the coherent control and incoherent  control for
quantum systems, hence it is easy to be generalized to
finite-dimensional systems. Finally, this analysis provides a new
control scheme, namely quantum states can be manipulated by probe
state-dependent coherent control on the quantum system.

Throughout this paper, we describe the state of the controlled
quantum system $s$ by a density matrix
$\rho_s$, a positive, unit trace operator on the Hilbert space  ${\cal H}_s$
of the system. The convex set of all possible states is
represented by ${\cal P}_s$ \cite{romano06}. Its boundary $\partial
{\cal P}_s$ is a set of pure states satisfying $\rho_s^2=\rho_s.$ By
the definition \cite{romano06}, the system $s$ is controllable if
and only if for all pairs $(\rho_i,\rho_f)\in {\cal P}_s\times {\cal
P}_s$, there exists a set of controls $\vec{g}$ such that
$\rho_s(t=0)=\rho_i$ and $\rho_s(t,\vec{g}=\vec{g}_{\mbox{\tiny
fixed}})=\rho_f$ for some $t\ge 0.$ Suppose that the quantum system
$s$ interacts with an initially unentangled probe $p$, whose density
matrix is denoted by $\rho_p$ on its Hilbert space ${\cal H}_p.$ The
time evolution of the composite system of the controlled system and
the probe is governed by $H=H_s+H_p+H_I,$ where $H_s$ and $H_p$ are
the free Hamiltonian for the controlled system and the probe,
respectively, while $H_I$ denotes the coupling of the system and the
probe. Since local transformations do not affect the controllability
of the system \cite{romano06}, $H_s$ and $H_p$ can be ignored safely
in the later analysis.

The paper is organized as follows. In Sec.\,{\rm II}, we study the
controllability of a two-dimensional system in the indirect control
scheme, then we extend the approach to arbitrary finite-dimensional
systems in Sec.\,{\rm III}. Conclusions and discussions are presented
in Sec.\,{\rm IV}.

\section{Two-dimensional systems}

Consider a two-dimensional system $s$ coupled to a probe $p$ modeled
as another two-dimensional system. The interaction Hamiltonian
describing such a composite system takes the form
\begin{equation}
H_I(\vec{g})=(g_1\sigma_{s}^{z}+g_2\sigma_{s}^{+}+g_{2}^{*}
\sigma_{s}^{-})\otimes(g_3\sigma_{p}^{x}+g_4\sigma_p^z),\label{eq1}
\end{equation}
where $\vec{g}=(g_1,g_2,g_3,g_4)$ are coupling constants,
$\sigma_s^{+,-,z}$ and $\sigma_p^{x,z}$ denote the Pauli matrices
acting on Hilbert space ${\cal H}_s$ and ${\cal H}_p$, respectively.

We now show that the two-dimensional system is controllable governed
by $H_I$.
The complete controllability requires that one can steer the system
$s$ from {\em arbitrary} initial states to an arbitrary pure or mixed target
state. This requirement on the initial states can be partly lifted by requiring that the
interaction Hamiltonian $H_I$ is unchanged up to $\vec{g}$ under the
local unitary  transformation
\begin{eqnarray}
F(\theta,\phi)&=&\left(
\begin{array}{cc}
\cos\frac{\theta}{2}  & -e^{i\phi}\sin\frac{\theta}{2} \\
e^{-i\phi}\sin\frac{\theta}{2}  & \cos\frac{\theta}{2}
\end{array}
\right)_s\otimes I_p\nonumber\\
&\equiv&f_s\otimes I_p,
\end{eqnarray}
where $0\le\phi\le 2\pi$, $0\le\theta\le\pi$ and $I_p$ is the
identity operator on ${\cal H}_p$. By unchanged we mean
$H_I(\vec{g{'}})=FH_I(\vec{g})F^{\dagger}$, namely the
transformation $F(\theta,\phi)$ changes the coupling constants (or
the controls) in the Hamiltonian only.   The proof is
straightforward. Suppose the composite system is initially prepared
on a state $\rho(0)=\rho_s(0)\otimes\rho_p(0)$.
 The final state of the quantum system $s$
reads ($F=F(\theta,\phi)$)
\begin{eqnarray}
\rho_s(t,\vec{g})&=& \mbox{Tr}_p(u\rho(0)u^{\dagger})\nonumber\\
&=& \mbox{Tr}_p(uF^{\dagger}F\rho(0)F^{\dagger}Fu^{\dagger})\nonumber\\
&=&f_s^{\dagger}\mbox{Tr}_p(u{'}\rho{'}(0)u{'}^{\dagger})f_s,
\label{proof1}
\end{eqnarray}
where the trace is taken over the probe,  $u=e^{-iH_I(\vec{g})t}$,
$u{'}=e^{-iH_I(\vec{g{'}})t}$, and $ \rho{'}(0)=F\rho(0)F^{\dagger}$
represents a set of states which have the same spectrums  as
$\rho(0)$. Since the two sets
$\{f_s\rho_s(t,\vec{g})f_s^{\dagger}\}$ and $\{\rho_s(t,\vec{g})\}$
are in one-to-one correspondence, thus  the quantum system $s$ is
controllable initially in $\{\mbox{Tr}_p\rho{'}(0)\}$ if and only if
it is controllable initially in $\{\mbox{Tr}_p\rho(0)\}$
\cite{note1}. As by proper choice of the unitary transformation
$f_s$, the initial state can be always transformed to the form
\begin{equation}
\rho_s(0)=\rho_i=p_s(0)|0\rangle_s\langle 0|
+(1-p_s(0))|1\rangle_s\langle 1|, \label{444}
\end{equation}
({\it where $|0\rangle_s$ and $|1\rangle_s$ denote the ground and
excited states of the quantum system, respectively, and $0\le
p_s(0)\le 1$}) the system is completely controllable if we can steer
the system to arbitrary target states from the initial state  in the
form (\ref{444}).

We then show that the Hamiltonian $H_I$ can
drive the quantum system from the initial state $\rho_s(0)$ to an
arbitrary final state.
The time evolution operator for the composite system governed by
$H_I$ can be written as
\begin{eqnarray}
U(t)=U_{+}^s(t)|+\rangle_p\langle +|+U_{-}^s(t)|-\rangle_p\langle
-|,
\end{eqnarray}
where
$|+\rangle_p=\cos\frac{\theta}{2}|1\rangle_p+\sin\frac{\theta}{2}|0\rangle_p$,
$|-\rangle_p=\sin\frac{\theta}{2}|1\rangle_p-\cos\frac{\theta}{2}|0\rangle_p$,
and  $U_{\pm}^s(t)$ satisfy
\begin{eqnarray}
i\hbar\frac{\partial}{\partial{t}}U_{\pm}^s(t)=H_{\pm}^sU_{\pm}^s(t),
\end{eqnarray}
with
\begin{eqnarray}
H_{\pm}^s&=&\pm\sqrt{g_3^2+g_4^2}
(g_1\sigma_{s}^{z}+g_2\sigma_{s}^{+}+g_{2}^{*}\sigma_{s}^{-}),\nonumber\\
\sin\theta&=&\frac{g_3}{\sqrt{g_3^2+g_4^2}}.\nonumber\\
\end{eqnarray}
We assume that the probe is initially prepared on the state
\begin{eqnarray}
\rho_p(0)=p_p(0)|0\rangle_p\langle 0|+(1-p_p(0))|1\rangle_p\langle
1|,
\end{eqnarray}
in terms of $|+\rangle_p$ and $|-\rangle_p$, $\rho_p(0)$ can be
rewritten as,
\begin{eqnarray}
 \rho_p(0)&=&\rho_p^{++}|+\rangle_p\langle +|+\rho_p^{--}|-\rangle_p\langle
-|\nonumber\\
&+&\rho_p^{+-}|+\rangle_p\langle -|+\rho_p^{-+}|-\rangle_p\langle +|
\end{eqnarray}
where $0\leq p_p(0)\leq 1$, and
$\rho_p^{++}=\cos^2\frac{\theta}{2}-p_p\cos\theta$,
$\rho_{p}^{--}=1-\rho_p^{++}$, $\rho_p^{+-}=\frac 12
\sin\theta-p_p\sin\theta$, as well as  $\rho_p^{-+}=\rho_p^{+-}$. We
shall discuss how to prepare such a state for the probe at the end
of this section. After some algebras, we obtain  the density matrix
for the quantum system at time $t$
\begin{eqnarray}
\rho_s(t)&=&p_s\rho_p^{++}U_+^s(t)|0\rangle_s\langle
0|U_+^{s\dagger}(t)\nonumber\\
&+&p_s\rho_p^{--}U_-^s(t)|0\rangle_s\langle
0|U_-^{s\dagger}(t)\nonumber\\
&+&(1-p_s)\rho_p^{++}U_+^s(t)|1\rangle_s\langle
1|U_+^{s\dagger}(t)\nonumber\\
&+&(1-p_s)\rho_p^{--}U_-^s(t)|1\rangle_s\langle 1|U_-^{s\dagger}(t),
 \label{fe1}
\end{eqnarray}
define
\begin{eqnarray}
 |\psi_{+0}\rangle=U_{+}^s|0\rangle_s, \ \
|\psi_{-0}\rangle=U_-^s|0\rangle_s,
\nonumber\\
|\psi_{-1}\rangle=U_-^s|1\rangle_s,\ \
|\psi_{+0}^{\bot}\rangle=U_{+}^s|1\rangle_s, \label{def}
\end{eqnarray}
clearly $|\psi_{+0}\rangle$ and $|\psi_{+0}^{\bot}\rangle$ are
normalized and orthogonal. Rewrite the density matrix for the
quantum system in the basis spanned by $\{ |\psi_{+0}\rangle,
|\psi_{+0}^{\bot}\rangle \}$, we obtain
\begin{eqnarray}
\rho_s(t)&=&\left(
\begin{array}{cc}
\rho_s^{00}(t)  & \rho_s^{01}(t) \\
\rho_s^{10}(t)  &\rho_s^{11}(t)
\end{array}%
\right),\label{density1}
\end{eqnarray}
where (setting $p_p(0)=p_p, $ and $p_s(0)=p_s$)
\begin{eqnarray}
\rho_s^{10}(t)&=&(\rho_s^{01})^*=\frac{\rho_p^{--}}{2}\sin 2\alpha
e^{-i\beta}(2p_s-1),\nonumber\\
\rho_s^{00}(t)&=&p_s\rho_p^{++}+(1-p_s)\rho_p^{--}\sin^2\alpha+p_s\rho_p^{--}\cos^2\alpha,\nonumber\\
\rho_s^{11}(t)&=&(1-p_s)\rho_p^{++}+p_s\rho_p^{--}\sin\alpha^2\nonumber\\
&+&(1-p_s)\rho_p^{--}\cos^2\alpha.
\nonumber\\
\end{eqnarray}
Here $\alpha$ and $\beta$ is defined by,
$$|\psi_{-0}\rangle=
\cos\alpha|\psi_{+0}\rangle+\sin\alpha
e^{i\beta}|\psi_{+0}^{\bot}\rangle,$$ equivalently,
$$|\psi_{-1}\rangle=
\sin\alpha e^{-i\beta}|\psi_{+0}\rangle+\cos\alpha
|\psi_{+0}^{\bot}\rangle.$$ Namely,
\begin{eqnarray}
\cos\alpha=|\langle \psi_{+0}|\psi_{-0}\rangle|,\nonumber\\
\beta=\mbox{Arg}\langle\psi_{+0}^{\bot}|\psi_{-0}\rangle
-\mbox{Arg}\langle\psi_{+0}|\psi_{-0}\rangle,
\end{eqnarray}
where  $\mbox{Arg}$  denotes a  function that extracts the angular
component (sometimes called the phase angle) of a complex number.

For a given target state, we need to determine the control
parameters $\vec{g}$ and initial state parameter $p_p$ from $\alpha$
and $\beta$ as well as the conditions that guarantee
$|\psi_{+0}\rangle$ is an arbitrary pure state of the system.
Dependence of $\vec{g}$ on $\alpha$ and $\beta$ is so complicated
that explicit expressions cannot be found for  general cases. We
will present an example to show how to calculate this dependence at
the end of this paragraph. Two observations can be made from
Eq.(\ref{def}). (1) $|\psi_{+0}\rangle$ can be prepared for the
quantum system to be an arbitrary pure state (unnormalized), this
can be done by control $g_1$ and $g_2$ in $H_{\pm}^s$. The reason is
that $|\psi_{+0}\rangle$ is a solution to the Schr\"odinger equation
$i\hbar\frac{\partial}{\partial
t}|\psi_{+0}\rangle=H_{+}^s|\psi_{+0}\rangle$ with the initial state
$|0\rangle_s$. Since $i\sigma_s^z \in su(2)$ and $i\sigma_s^{\pm}
\in su(2)$ are generators of the Lie algebra $SU(2)$,
$|\psi_{+0}\rangle$ then can be manipulated to be any pure state in
$\partial {\cal P}_s$ by varying $g_1$ and $g_2$; (2) $\alpha$ (or
$\beta$) are controllable by $g_3$ and $g_4$ resulting in the change
of the overlap between  $|\psi_{-0}\rangle$ and $|\psi_{+0}\rangle.$
As a result of these observations, we conclude that $\rho_s^{10}(t)$
can be manipulated to be zero for any $p_s$ and $p_p,$ leading to
\begin{equation}
\rho_s(t)=\rho_s^{00}(t)|\psi_{+0}\rangle\langle
\psi_{+0}|+\rho_s^{11}(t)|\psi_{+0}^{\bot}\rangle\langle
\psi_{+0}^{\bot}|.
\end{equation}
Obviously, $(\rho_s^{00}+\rho_s^{11})=1,$ and $\rho_s^{00}$ can be
arbitrarily  controlled in the interval $[0,1]$ by changing $p_p$.
Therefore, the quantum system is controllable with $H_I$. To be
specific, from the condition $\rho_s^{10}(t)=0$ we can easily find
that
\begin{eqnarray}
\cos\theta&=&\frac{1}{1-2p_p},\nonumber\\
\alpha&=&n\pi, n=0,\pm 1,\pm 2,...,\nonumber\\
\label{spe2}
\end{eqnarray}
Suppose that the target state is $\rho_s^{00}(t)=q,  (0\leq q \leq
1)$, we find
\begin{equation}
p_p=\frac{q-p_s\cos^2\frac{\theta}{2}
-\sin^2\frac{\theta}{2}\sin^2\alpha-p_s\sin^2\frac{\theta}{2}\cos
2\alpha}{\cos\theta\sin^2\alpha+p_s\cos\theta\cos
2\alpha-p_s\cos\theta}. \label{spe1}
\end{equation}
Eqs.(\ref{spe2}) and (\ref{spe1}) together determine the parameters
$g_3$, $g_4$ and $p_p$.

In comparison with earlier work \cite{romano06,romano06l}, it seems
that our proposal needs more real parameters to control a two-level
system. This is not true. In fact, $\rho_s^{10}(t)=0$ is not
necessary to prove the controllability.  The reason is as follows.
Diagonalizing the density matrix Eq.(\ref{density1}), we can write
the state of the quantum system as
\begin{equation}
\rho_s(t)=E_+(t)|\psi_+(t)\rangle\langle\psi_+(t)|
+E_-(t)|\psi_-(t)\rangle\langle\psi_-(t)|,
\end{equation}
with the eigenvalues and the corresponding eigenvectors as follows
\begin{eqnarray}
&&E_{\pm}(t)=\frac 1 4\left[(\rho^{00}+\rho^{11})\pm
\sqrt{(\rho^{00}-\rho^{11})^2+4|\rho^{01}|^2}\right], \nonumber \\
&&|\psi_+(t)\rangle=\cos\frac{\Gamma}{2}|\psi_{+0}\rangle+
\sin\frac{\Gamma}{2}e^{i\gamma}|\psi_{+0}^{\bot}\rangle, \nonumber
\end{eqnarray}
where $|\psi_-(t)\rangle$ is orthogonal to $|\psi_+(t)\rangle$,
$\gamma=\mbox{Arg}(\rho^{01})$, and the mixing angle $\Gamma$ are
defined by $\cos\Gamma=
(\rho^{00}-\rho^{11})\left[(\rho^{00}-\rho^{11})^2+4|\rho^{01}|^2\right]^{1/2}$.
Note that $|\psi_+(t)\rangle$ (or $|\psi_-(t)\rangle$) can be
controlled to be an arbitrary pure state regardless of $\Gamma$ and
$\gamma$, since $|\psi_{+0}\rangle$ is controllable (in the sense of
pure states). Hence, two real parameters in $H$ are enough to
control $|\psi_+(t)\rangle$. This together with the parameter to
control $E_+$ (from zero to $1/2$) \cite{note2}, only three real
parameters are required to control the two-level system.

\begin{figure}
\includegraphics*[width=0.8\columnwidth,height=0.7\columnwidth]{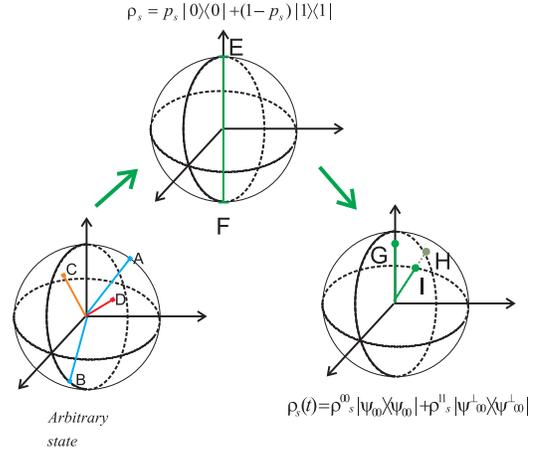}
\caption{(Color online) Schematic illustration of the proof  for
controllability. A, B, C and D in the left Bloch ball represent
arbitrary pure and mixed states, in which A and B denote arbitrary
pure states, while C and D denote arbitrary mixed states. The
controllability requires that the quantum system can be controlled
from A to B and C to D,  from C  to A, B and D as well as from A  to
C, B and D. We first showed that it is not necessary to require the
quantum system to be controllable starting from an arbitrary initial
state (i.e., an arbitrary point in the Bloch ball), provided the
interacting Hamiltonian $H_I$ is unchanged under $F(\theta,\phi)$,
rather it is enough to control the system initially from a state on
the green line in the middle of Bloch ball. The actual control can
be done by first rotating the initial state G determined by the
target state to H, and then changing the purity of the state G to
reach the target state I.} \label{fig1}
\end{figure}

These discussions implies that a two-level system can be
controlled by the Hamiltonian Eq.(\ref{eq1}) together with the
initial state preparation of the probe. The probe is not required to
be prepared in an arbitrary initial state, rather it is a specified
state determined according to the target state. In
fact, this requirement on the probe can be removed by adding a
control in the Hamiltonian (see Eq.(\ref{spe1})). This holds true
for arbitrary finite dimensional systems we will discuss
in the next section.

Before closing this section, we propose a possible way to prepare the probe
on the required mixed state, $\rho_p(0)=p_p(0)|0\rangle_p\langle
0|+(1-p_p(0))|1\rangle_p\langle 1|$. For this purpose, we place the
probe in a thermal environment at temperature $T$.  At equilibrium,
the density matrix of the probe would take the above form with
$p_p=e^{-\beta E_0}/(e^{-\beta E_0}+e^{-\beta E_1})$, where
$\beta=1/k_BT$, and $E_0$($E_1$) is the energy corresponding to
state $|0\rangle$ ($|1\rangle$). By varying the energy spacing
$|E_1-E_0|$, we can in principle obtain the required $p_p$. Note
that $p_p$ is not required to vary between 0 and 1 \cite{note2}, we
conclude that the mixed state prepared in this way meets requirement
in this proposal.

\section{Finite-dimensional system}

In order to extend the approach to finite-dimensional systems, we
recall that the controllability in this case can be expressed in the
following way. For any state $\rho_s$ in ${\cal P}_s$, the quantum
system is controllable if and only if there exists a Kraus map
$\Phi_{\rho_s}$ such that $\Phi_{\rho_s}(\rho_s)=\rho_f$ for all
state $\rho_s$ in ${\cal P}_s.$ In this section, we shall show that
the following Hamiltonian
\begin{eqnarray}
H_I&=&\left(\sum_{j\neq i=1}^Ng_{ij}|i\rangle_s\langle
j|\right)\otimes \left(\sum_{m\neq n}^N f_{mn}|m\rangle_p\langle
n|\right)\nonumber\\
&=&h_s\otimes h_p,
\end{eqnarray}
which describes interaction between the system $s$ and the probe
$p$, would give rise to the Kraus map $\Phi_{\rho_s}$ required for
the controllability. Here the probe $p$ is modeled as a finite
dimensional system with the same dimension as the system.
Assume that the composite system is initially
on an uncorrelated state $\rho(0)=\rho_s(0)\otimes\rho_p(0)$, and
the target state of the quantum system reads
\begin{eqnarray}
\rho_s(t)=Tr_p[{\cal U}(t)\rho_s(0)\otimes\rho_p(0){\cal
U}^\dagger(t)],
\end{eqnarray}
where ${\cal U}(t)$ denotes the time evolution operator of the whole
system.  Preparing initially the probe on state ($\sum_n P_n=1$)
\begin{eqnarray}
\rho_p(t=0)&=&\sum_nP_n\ket{n}_p\bra{n}\nonumber\\
&=&\sum_{M,N}p_{MN}|M\rangle_p\langle N|,
\end{eqnarray}
we obtain
\begin{equation}
\rho_s(t)=\sum_{M}p_{MM}U_M(t)^s\rho_s(0)U_M^{s\dagger}(t).\label{fs1}
\end{equation}
Here, $|M\rangle_p$ is an eigenstate of $h_p$, i.e.,
$h_p|M\rangle_p=E_p|M\rangle_p$.  It is easy to find that $U_M^s(t)$
satisfies
\begin{eqnarray}
i\hbar\frac{\partial U_{M}^s(t)}{\partial
t}&=&H_{M}^sU_{M}^s(t),\nonumber \\
H_{M}^s&=&E_M\sum_{i\neq j=1}^{N}g_{ij}\ket{i}_s\bra{j}.
\end{eqnarray}
The above results can be readily derived by writing ${\cal
U}(t)=\sum_M U_M^s(t)|M\rangle_p \langle M|,$ and $\rho_s(t)={\mbox
Tr}{\cal U}(t)\rho_p(0)\otimes\rho_s(0){\cal U}^{\dagger}(t).$ With
these results,
  the Kraus map $\Phi_{\rho_s}$  is
\begin{equation}
\Phi_{\rho_s}(\rho_s(0))=\sum_{M}K_{M}(t)\rho_s(0)K^{\dag}_{M}(t),
\end{equation}
where $K_{M}(t)=\sqrt{p_{MM}} \,U_{M}^s(t)$. It has been proved that
$U_{M}^s(t)$ can drive all initial pure states of the quantum system
$s$ into an arbitrary  pure state at some time $t\ge 0$
\cite{fu01,schirmer01}. So  $U_{M}^s(t)$ can be written as
\begin{equation}
U_{M}=|\phi_M(t)\rangle\langle\varphi_M(0)|,
\end{equation}
where both $|\phi_M(t)\rangle$ and $|\varphi_M(0)\rangle$ are
arbitrary and in $\partial {\cal P}_s$. This together with the
hypothesis that the probe can be initially prepared in an arbitrary
state, show controllability of the  quantum system $s$. In contrast
to the earlier study \cite{wu07}, in which the authors presented a
constructive proof of complete kinematic state controllability for
finite-dimensional open systems via the Kraus map, we here not only
provide the details to realize such a Kraus map but also derive the
equations for parameters to control the system from an arbitrary
initial state to an arbitrary target state. It is worth pointing out
that the realization of the Kraus map is not a trivial task, many
works in literature are devoted to this task,  in particular not all
Kraus maps allowed in quantum mechanics can be used to control a
quantum system. This problem  was shown in Ref.\,\cite{altafini04}
and  has been realized in Ref.\,\cite{pechen06}.

Now we present a general formulism to show
that the indirect control can be represented as a combination of
coherent control for the quantum system $s$ and preparation of the
initial state of the probe $p$. Consider an initial state of the
composite system $\rho(0)=\rho_s(0)\otimes\rho_p(0)$, and a
Hamiltonian $H=H_s\otimes H_p$ that governs the evolution of the
composite system. Suppose that both the quantum system and the probe
are $N$-dimensional, and the quantum system is coherently
controllable driven by $H_s$. The latter assumption means $H_s$ can
drive the quantum system from an arbitrary pure state to any other
pure state,
i.e.,
$\{\, |\phi_i\rangle_s \equiv
e^{-iH_st}|i\rangle_s \,|\, i=1,2,...,N\}$
spans a basis for the quantum system, and $|\phi_i\rangle_s$
is an arbitrary pure state in $\partial {\cal P}_s$ for any $i$. The
density matrix for the quantum system at time $t$ reads
\[
\rho_s(t)=\sum_{m}  \ _p \langle m |e^{-i(H_s\otimes H_p)
t}\rho_s(0)\otimes\rho_p(0)e^{i(H_s\otimes H_p) t}|m\rangle_p.
\]
Choosing $H_p|m\rangle_p=E_m|m\rangle_p$, we obtain
\begin{equation}
\rho_s(t)=\sum_jp_j\sum_m
\mbox{$_p\langle$}m|\rho_p(0)|m\rangle_p|\phi_j(E_mt)\rangle\langle
\phi_j(E_mt)|,
\end{equation}
where we have set the initial state of the system as
$\rho_s(0)=\sum_jp_j|j\rangle_s\langle j|$, and
used the relation $|\phi_j(E_mt)\rangle=e^{-iH_sE_mt}|j\rangle_s$.
Suppose the target state is ($\sum_j q_j=1$)
\begin{equation}
\rho_s(T)=\sum_m^N q_m|\phi_m(T)\rangle\langle \phi_m(T)|,
\end{equation}
we find the condition to determine the controls in $H_s$ and $H_p$
\begin{eqnarray}
&& q_{\alpha}=\sum_jp_j\sum_m
\mbox{$_p\langle$}m|\rho_p(0)|m\rangle_p|c_{\alpha}^{jm}|^2,\nonumber\\
&& \sum_jp_j \sum_m
\mbox{$_p\langle$}m|\rho_p(0)|m\rangle_pc_{\beta}^{jm}
(c_{\gamma}^{jm})^*=0,\nonumber\\
&& \ \ \ \ \ \ \alpha, \beta, \gamma=1,2,...,N,
\end{eqnarray}
where $c_{\alpha}^{jm}$ is defined by $|\phi_j(E_m
t)\rangle=\sum_{\alpha} c_{\alpha}^{jm}|\phi_{\alpha}(T)\rangle.$
 These results suggest  that the manipulation of a quantum
system may be realized by  coherent controls for the quantum system
conditioned on the spectrum of the initial density matrix of the
probe.

\section{conclusion}

In this paper we have presented a scheme for indirect control of a
quantum system coupled to a probe. This control scheme is actually a
combination of coherent control conditioned on the initial state of
the probe. For a two-level system, we have proved that the
restriction on the initial state for the quantum system can be
partly removed. This simplifies the formulation for the
controllability. The scheme has been generalized to arbitrary finite
dimensional systems and equations to determine the controls  are
given. We would like to note that the probe is not required to be
initially prepared in an arbitrary state but with a specified
spectrum. This requirement can be lifted by adding more controls in
the composite systems.

This work was supported by  NSF of China under grant No. 10775023
and 10675085.


\begin{references}
\bibitem{note1}
This also can be understood as follows. Consider the Schr\"odinger
equation
$i\hbar\frac{\partial}{\partial{t}}|\varphi(t)\rangle=H_I|\varphi(t)\rangle,
$ where $|\varphi(t)\rangle$ is the wavefunction of the composite
system. By the time-independent transformation $F(\theta,\phi)$ ,
$|\varphi(t)\rangle\rightarrow{F^{\dagger}|\phi{'}(t)\rangle}$  we
find $
i\hbar\frac{\partial}{\partial{t}}(F^{\dagger}|\phi{'}(t)\rangle)=
H_I(F^{\dagger}|\phi{'}(t)\rangle),$ and
$i\hbar\frac{\partial}{\partial{t}}|\phi{'}(t)\rangle=
FH_IF^{\dagger}|\phi{'}(t)\rangle.$ Since
$H_I(\vec{g}{'})=FH_I(\vec{g})F^{\dagger}$, we claim that there
exists a one-to-one correspondence in sets
$\{Tr_p(|\phi{'}(t)\rangle\langle\phi{'}(t)|)\}$ and
$\{Tr_p(|\phi(t)\rangle\langle\phi(t)|\}.$ Therefore, if
$\{Tr_p(|\phi{'}(t)\rangle\langle\phi{'}(t)|)\}$ covers all states
in ${\cal P}_s$, $\{\mbox{Tr}_p(|\phi(t)\rangle\langle\phi(t)|\}$ is
a convex set of all possible states for the two dimensional system.
\bibitem{note2} We can obtain the state with $E_+ \in (1/2,2]$ by
controlling $|\psi_+(t)\rangle$, because of $\langle
\psi_+(t)|\psi_-(t)\rangle=0.$
\end{references}
\end{document}